\documentstyle[12pt,cite]{article}

\begin{document}

\begin{center}

{\bf Neutrino Oscillations in Moving and Polarized Matter 
under the Influence of Electromagnetic Fields}

{\bf A.E. Lobanov,
A.I. Studenikin\footnote{\normalsize E-mail: studenik@srdlan.npi.msu.su}}


{\bf Department of Theoretical Physics, Moscow State University,
119899 Moscow, Russia}

{\it Abstract}
\end{center}
{\it Within the recently proposed \cite {ELSt99,
ELStpl00, ELSt00} Lorentz invariant formalism for description of neutrino spin
evolution in presence of an arbitrary electromagnetic fields
matter motion and polarization effects are considered. It is shown that in 
the case of matter moving with relativistic speed parallel to neutrino 
propagation, matter effects in neutrino spin (and also flavour) 
oscillations are suppressed. In the case of relativistic motion of matter 
in the opposite direction in respect to neutrino propagation, sufficient 
increase of effects of matter in neutrino oscillations is predicted.
These phenomena could have important consequences in 
different astrophysical environments.}


In \cite {ELSt99, ELStpl00, ELSt00}
the Lorentz invariant formalism for neutrino motion in
non-moving and isotropic matter
under the influence of an arbitrary configuration of electromagnetic fields
have been developed.
We have derived the neutrino spin evolution Hamiltonian that accounts
not only for the transversal to the neutrino momentum components of
electromagnetic field but also for the longitudinal components. With
the using of the proposed Hamiltonian it is possible to consider
neutrino spin precession in an arbitrary configuration of
electromagnetic fields including those that contain strong
longitudinal components. We have also considered the new types
of resonances in the neutrino spin precession
$\nu_{L}\leftrightarrow\nu_{R}$ that could appear when neutrinos propagate
in matter under the influence of different electromagnetic field
configurations. Within the proposed approach the parametric resonance 
of neutrino oscillations in electromagnetic wave field with periodically
varying time-dependent amplitude has been also studied \cite {DSthp0102099}.

In the studies \cite {ELSt99,ELStpl00, ELSt00} of the neutrino spin evolution
we have focused mainly on description of influence of different
electromagnetic fields, while modelling the matter we confined
ourselves to the most simple case of non-moving and unpolarised
matter.  Now we should like to go further ( see also \cite {DELSt01})
and to generalize our
approach for the case of moving and polarized homogeneous matter.
In this paper we develop the covariant description of neutrino oscillations
in moving and polarized matter under the influence of electromagnetic fields. 
This approach is valid for accounting of matter motion and polarization for 
arbitrary speed of matter. It is shown for the first time that in the case of 
relativistic motion of matter, the value of effects of matter in neutrino 
oscillations (spin oscillations as well as flavour oscillations) sufficiently 
depends on direction of matter motion in respect to neutrino propagation.

It should be noted here that effects of matter polarization in 
neutrino oscillations were considered previously in several papers
( see, for example, \cite {NSSV97,BGN99} and references therein). 
However, the used in refs. \cite {NSSV97,BGN99} procedure of
accounting for the matter polarization effect does not enable one to study 
the case of matter motion with relativistic speed. 
Within our approach we can reproduce corresponding results of 
\cite {NSSV97,BGN99} in the case of matter which is slowly moving or 
is at rest.

To derive the equation for the neutrino
spin evolution in electromagnetic field $F_{\mu \nu}$ in moving and 
polarized matter we
again start from the
Bargmann-Michel-Telegdi (BMT) equation \cite{BMT59} for the spin vector
$S^{\mu}$ of a neutral particle that has the following form
\begin{equation}
{dS^{\mu} \over d\tau} =2\mu \big\{ F^{\mu\nu}S_{\nu} -u^{\mu}(
u_{\nu}F^{\nu\lambda}S_{\lambda} ) \big\} +2\epsilon \big\{ {\tilde
F}^{\mu\nu}S_{\nu} -u^{\mu}(u_{\nu}{\tilde
F}^{\nu\lambda}S_{\lambda}) \big\},
\label{1}
\end{equation}
This form of the BMT equation corresponds to the case of the particle moving
with constant speed,
$\vec \beta=const$, ($u_\mu=(\gamma,\gamma \vec \beta),
\gamma=(1-\beta^2)^{-1/2})$, in presence of an electromagnetic
field $F_{\mu\nu}$.
Here $\mu$ is the fermion magnetic moment and
${\tilde F}_{\mu\nu}$ is the dual electromagnetic field tensor.
The neutrino spin vector satisfies the usual conditions,
$S^2=-1$ and $S^{\mu}u_\mu =0$.
Equation (1) covers also the case of a neutral fermion having
static non-vanishing electric dipole moment, $\epsilon$.  Note that
the term proportional to $\epsilon$ violates $T$ invariance.

The BMT spin evolution equation (1) is derived in the frame of
electrodynamics, the model which is $P$ invariant.
Our aim is to generalize this equation for the case when effects of
various neutrino interactions (for example, weak
interaction for which $P$ invariance is broken)
with moving and polarized matter are also taken into account.
Effects of possible $P$ nonconservation and nontrivial properties of matter
(i.e., its motion and polarization) have to be reflected in the equation that
describes the neutrino spin evolution in an electromagnetic field.

The Lorentz invariant
generalization of eq.(1) for our case can be obtained by the substitution
of the electromagnetic field tensor
$F_{\mu\nu}=(\vec E,\vec B)$ in the following way:
\begin{equation}
F_{\mu\nu}\rightarrow F_{\mu\nu}+G_{\mu\nu}.
\label{2}
\end{equation}

In evaluation of the tensor $G_{\mu \nu}$ we demand that the neutrino
evolution
equation has to be linear over
the neutrino spin vector $S_{\mu}$, electromagnetic field $F_{\mu\nu}$,
and such characteristics of matter (which is composed of different fermions,
$f=e,\ n,\ p...$) as fermions currents
\begin{equation}
j_{f}^\mu=(n_f,n_f\vec v_f),
\label{3}
\end{equation}
and fermions polarizations
\begin{equation}
\lambda^{\mu}_f =\Big(n_f (\vec \zeta_f \vec v_f ),
n_f \vec \zeta _f \sqrt{1-v_f^2}+
{{n_f \vec v_f (\vec \zeta_f \vec v_f )} \over {1+\sqrt{1-v_f^2}}}\Big).
\label{4}
\end{equation}
Here $n_f$, $\vec v_f$, and $\vec \zeta_f \
(0\leq |\vec \zeta_f |^2 \leq 1)$ denote, respectively,
the number density of the background fermions $f$, the
speed of the reference frame in which the mean
momentum of fermions $f$ is zero, and the mean value of the polarization
vectors of the background fermions $f$ in the above mentioned reference frame.
Note that, as it follows from eq.(4), if a component $f$ 
of matter is not moving, $\vec v_f =0$, the fermion $f$  polarization is 
expressed as,
\begin{equation}
\lambda^{\mu}_f =\big(0,n_f \vec \zeta _f \big).
\label{5}
\end{equation}

The mean value of the background fermion $f$ polarization vector, 
$\vec \zeta_f$, 
is determined by the two-step averaging procedure. The polarization 
of the moving 
fermion is described by the relativistic generalization of the spin operator:
\begin{equation}
\vec O=\gamma _{0} \vec \Sigma - \gamma _{5} {\vec p \over p_{0}} - 
\gamma _{0} {{\vec p \big(\vec p \vec \Sigma \big)} \over 
{p_{0} (p_{0} + m)}}, 
\label{6}
\end{equation}
where
$$
\vec \Sigma = \Bigg(\matrix{\vec \sigma & 0 \cr
                       0 & \vec \sigma \cr }\Bigg),$$ 
$\vec \sigma$ is the Pauli matrices
and $\vec p, p_{0}, m$ are, respectively, the fermion momentum, energy and 
mass. At the first step the average of the fermion spin operator $\vec O$
have to be evaluated over the fermion quantum state in a given 
electromagnetic field,
\begin{equation}
<\vec O >= \int \limits_{}^{} \Psi_{f}^{+} (x) \vec O \Psi_{f}(x) dx,
\label{7}
\end{equation}
here $\Psi_{f}(x)$ is the exact solution of the Dirac equation for the fermion 
accounting for the external electromagnetic field. 

The second-step averaging is performed over the fermion statistical 
distribution density function, $\rho_{f}(\{ n \})$:
\begin{equation}
\vec \zeta _{f} = { \sum\limits_{\{n\}}^{} <\vec O> \rho_{f} (\{n\}) \over
\sum\limits_{\{n\}}^{} \rho_{f} (\{n\})}.
\label{8}
\end{equation}
In the case when the background fermions $f$ can be described as an ideal gas, 
$\rho_{f}(\{n\})$ is nothing but the Fermi-Dirac distribution function.

For each type of the fermions $f$ there
are only three vectors, $u^{f}_\mu, \ j^{f}_\mu,$ and $\lambda ^{f}_\mu$,
using which the tensor $G_{\mu \nu}$ have to be constructed.
If $j^{f}_\mu$ and $\lambda^{f}_{\mu}$ are slowly varying functions
in space and time
(this condition is similar to one imposed on the electromagnetic
field tensor $F_{\mu \nu}$ in the derivation of the BMT equation)
then one can construct only four tensors (for each of the fermions
$f$) linear in respect to the characteristics of matter:
\begin{equation}
G_{1}^{\mu \nu}=\epsilon ^{\mu \nu \rho \lambda}u_{\lambda}j_{\rho}, \
\ \ G_{2}^{\mu \nu}=\epsilon ^{\mu \nu \rho \lambda}u_{\lambda}\lambda_{\rho},
\label{5}
\end{equation}
\begin{equation}
G_{3}^{\mu \nu}=u^{\mu}j^{\nu}-j^{\mu}u^{\nu}, \ \ \
G_{4}^{\mu \nu}=u^{\mu}\lambda ^{\nu}-\lambda ^{\mu}u^{\nu}.
\label{6}
\end{equation}
Thus,
in general case of neutrino interaction with different background
fermions $f$ we introduce for description of matter effects
antisymmetric tensor
\begin{equation}
G^{\mu \nu}= \epsilon ^{\mu \nu \rho \lambda}
g^{(1)}_{\rho}u_{\lambda}- (g^{(2)\mu}u^\nu-u^\mu g^{(2)\nu}),
\label{7}
\end{equation}
where
\begin{equation}
g^{(1)\mu}=\sum_{f}^{} \rho ^{(1)}_f j_{f}^\mu
+\rho ^{(2)}_f \lambda _{f}^{\mu}, \ \
g^{(2)\mu}=\sum_{f}^{} \xi ^{(1)}_f j_{f}^\mu
+\xi ^{(2)}_f \lambda _{f}^{\mu}.
\label{8}
\end{equation}
Summation is performed over the fermions $f$ of the background. The explicit
expressions for the coefficients $\rho_{f}^{(1),(2)}$ and $\xi_{f}^{(1),(2)}$
could be found if the particular
model of neutrino interaction is chosen.
In the usual notations the antisymmetric tensor $G_{\mu \nu}$ can be
written in the form,
\begin{equation}
G_{\mu \nu}= \big(-\vec P,\ \vec M),
\label{9}
\end{equation}
where
\begin{equation}
\vec M= \gamma \big\{(g^{(1)}_0 \vec \beta-\vec g^{(1)})
- [\vec \beta \times \vec g^{(2)}]\big\}, \
\vec P=- \gamma \big\{(g^{(2)}_0 \vec \beta-\vec g^{(2)})
+ [\vec \beta \times \vec g^{(1)}]\big\}.
\label{10}
\end{equation}
It worth to note that the substitution (2) implies that the magnetic $\vec B$
and electric $\vec E$ fields are shifted by the vectors $\vec M$ and $\vec P$,
respectively:
\begin{equation}
\vec B \rightarrow \vec B +\vec M, \ \ \vec E \rightarrow \vec E -\vec P.
\label{11}
\end{equation}

In the case of non-moving, $\vec v_f=0$, and unpolarized, $\vec \zeta _f=0,$
matter we get, in agreement with our previous result \cite {ELSt99, ELStpl00, 
ELSt00},
\begin{equation}
G_{\mu \nu}=\Big(\gamma \vec \beta\sum_{f}^{}\xi^{(1)}_f n_f ,
\gamma \vec \beta\sum_{f}^{}\rho^{(1)}_f n_f \Big).
\label{12}
\end{equation}

We finally
arrive to the following equation for the
evolution of the three-di\-men\-sio\-nal neutrino spin vector $\vec S
$ accounting for the direct neutrino interaction with electromagnetic
field $F_{\mu \nu}$ and matter (which is described by the tensor
$G_{\mu \nu}$):
\begin{equation}
{d\vec S \over dt}={2\mu \over \gamma} \Big[
{\vec S \times ({\vec B_0}+\vec M_0) \Big]+{2\epsilon \over \gamma}
\Big[{\vec S} \times (\vec E_0-\vec P_0)} \Big].
\label{13}
\end{equation}

The derivative in the left-hand
side of eq.(17) is taken with respect to time $t$ in the
laboratory frame, whereas the values $\vec B_0$ and $\vec E_0$ are
the magnetic and electric fields in the neutrino rest frame given in terms
of the transversal in respect to the
neutrino motion ($\vec F_{\perp}$) and longitudinal ($\vec F_{\parallel}$) 
fields $\vec F= \ \vec B,\vec E$ in the
laboratory frame,
\begin{equation}
\begin{array}{c}
\vec B_0=\gamma\big(\vec B_{\perp}
+{1 \over \gamma} \vec B_{\parallel} + \sqrt{1-{1 \over
\gamma^2}}\big[{\vec E_{\perp} \times \vec n}\big]\big),\\
\vec E_0=\gamma\big(\vec E_{\perp} +{1 \over \gamma} \vec E_{\parallel} -
\sqrt{1-{1 \over \gamma^2}}\big[{\vec B_{\perp} \times \vec
n}\big]\big), \vec n={\vec \beta}/\beta.
\label{14}
\end{array}
\end{equation}
The influence of matter on the neutrino spin evolution in eq.(17) is given by
the vectors $\vec M_0$ and $\vec P_0$ which in the rest frame of neutrino
can be expressed in terms of quantities determined in the laboratory
frame
\begin{equation}
\vec M_0=\gamma \vec \beta
\Big(g^{(1)}_0-{{\vec \beta \vec g^{(1)}} \over {1+\gamma ^{-1}}}\Big)
-\vec g^{(1)},
\label{15}
\end{equation}
\begin{equation}
\vec P_0=-\gamma \vec \beta
\Big(g^{(2)}_0-{{\vec \beta \vec g^{(2)}} \over {1+\gamma ^{-1}}}\Big)
+\vec g^{(2)}.
\label{16}
\end{equation}

Let us determine the coefficients $\rho^{(i)}_f$ and $\xi^{(i)}_f$
in eq.(12) for the particular
case of the electron neutrino propagation in moving and polarized
electron gas. We consider the standard model of interaction supplied with
$SU(2)$-singlet right-handed neutrino $\nu_R$. The neutrino effective
interaction Lagrangian reads
\begin{equation}
L_{eff}=-f^\mu \Big(\bar \nu \gamma_\mu {1+\gamma^5 \over 2} \nu \Big),
\label{17}
\end{equation}
where
\begin{equation}
f^\mu={G_F \over \sqrt2}\Big((1+4\sin^2 \theta _W) j^\mu_e -
\lambda ^\mu _e\Big).
\label{18}
\end{equation}
In this case neutrino electric dipole moment vanishes, $\epsilon
=0$, so that the coefficients $\xi^{(i)}_e =0$, and from the obvious
relation, $f_\mu=2\mu g_\mu^{(1)}$, it follows
\begin{equation}
\rho^{(1)}_e={G_F \over
{2\mu \sqrt2}}(1+4\sin^2 \theta _W), \ \rho^{(2)}_e=-{G_F \over {2\mu
\sqrt2}}.
\label{19}
\end{equation}
If for the neutrino magnetic moment we take the
vacuum one-loop contribution \cite {LeeShr77, Fu}
$$\mu_{\nu}={3
\over {8\sqrt2 \pi^2}}eG_F m_\nu,$$
then
$$\rho^{(1)}={4\pi^2\over 3em_\nu}(1+4\sin^2\theta _W),\ \
\rho^{(2)}=-{4\pi^2\over 3em_\nu}.$$

We should like to note that solutions of the derived eq.(17)
for the neutrino spin evolution in moving and polarized matter and,
correspondingly, the neutrino oscillation probabilities and effective
mixing angles $\theta_{eff}$ can be obtained for different
configurations of electromagnetic fields in a way similar to that
described in \cite{ELSt99, ELStpl00, ELSt00}.

Consider the case of neutrino propagating in the relativistic
flux of electrons. Using expressions for the vector $\vec M_0$, eqs.
(12), (19), we find,

\begin{equation}
\begin{array}{c}
{\vec {M}_0}=n_e\gamma{\vec\beta}\Big\{
\big(
\rho^{(1)}+\rho^{(2)}{\vec\zeta_{e}}{\vec v}_e
\big)
(
1-{\vec\beta}{\vec v}_e
)+\\
+\rho^{(2)}\sqrt{1-v^2_e}
\Big[
{
(\vec \zeta_{e}{\vec v}_e)(\vec\beta{\vec v}_e)
\over
1+\sqrt{1-v^2_e}
}-\vec\zeta_{e}\vec\beta
\Big]
+O(\gamma^{-1})
\Big\}.
\end{array}
\label{M_0} \end{equation}
In the case of slowly moving matter, $v_e\ll1$, we get
\begin{equation}
{\vec M}_0=n_e\gamma{\vec\beta}\Big(
\rho^{(1)}-\rho^{(2)}\vec\zeta_{e}\vec\beta \Big).
\label{21}
\end{equation}
For the unpolarized matter eq.(\ref{21}) reproduces the
Wolfenstein matter term \cite {Wolf78} and confirms our previous result
\cite{ELSt99,ELStpl00,ELSt00}.  In the opposite case of relativistic flux,
$v_e\sim 1$, we find,
\begin{equation}
{\vec M}_0=n_e\gamma{\vec\beta}
\Big(
\rho^{(1)}+\rho^{(2)}\vec\zeta_{e}{\vec v}_e
\Big)
\Big(
1-\vec\beta{\vec v}_e
\Big).
\end{equation}
If we
introduce the invariant electron number density, 
\begin{equation}
\begin{array}{c}
n_0=n_e\sqrt{1-v^2_e},
\end{array}
\end{equation}
then it follows,
\begin{equation}
\begin{array}{c}
{\vec M}_0=n_o\gamma{\vec\beta}
{(1-\vec\beta{\vec v}_e)\over\sqrt{1-v^2_e}}\Big(
\rho^{(1)}+\rho^{(2)}\vec\zeta_{e}{\vec v}_e
\Big).
\end{array}
\end{equation}
Thus, in the case of the parallel motion of neutrinos
and electrons of the flux, the matter effect contribution to the neutrino
spin evolution equation (\ref{13}) is suppressed. In the case of neutrino 
and matter relativistic motion ($\beta$ and $v_e \sim 1$) 
in opposite directions, the matter term $\vec M_0$ gets its maximum 
value
\begin{equation}
\vec M_{0}^{max}=2n_e\gamma{\vec\beta}
\Big(
\rho^{(1)}+\rho^{(2)}\vec\zeta_{e}{\vec v}_e
\Big),
\end{equation}
which is equal to the matter term 
derived for the case of slowly moving $(v_e \ll 1)$ matter times 
a factor ${2 \over \sqrt{1-v_{e}^2}}$. Therefor we predict sufficient
increase of matter effects in neutrino oscillations for neutrino  
propagating against the relativistic flux of matter.

Finally, let us consider neutrino spin oscillations in an arbitrary constant
magnetic field, $\vec B=\vec B_{\parallel} +\vec B_{\perp}$, and moving matter.
In the adiabatic approximation for the particular case of electron neutrinos
$\nu_e$ propagating in matter composed of electrons, $f=e$, the probability
of conversion $\nu_L \rightarrow \nu_R$ can be written in the form,
\begin{equation}
P_{\nu_L \rightarrow \nu_R} (x)=\sin^{2} 2\theta_{eff} \sin^{2}{\pi x 
\over L_{eff}},\end{equation}
\begin{equation}
sin^{2} 2\theta_{eff}={E^2_{eff} \over {E^{2}_{eff}+\Delta^{2}_{eff}}},
L_{eff}={2\pi \over \sqrt{E^{2}_{eff}+\Delta^{2}_{eff}}},
\end{equation}
where $E_{eff}=2\mu B_{\perp}$ (terms $\sim O(\gamma^{-1})$ are ommited  here),
and 
\begin{equation}
\Delta_{eff}= V(1-\vec \beta \vec v_{e}) +{2\mu B_{\parallel} \over \gamma},
V={G_F \over \sqrt{2}}n_e.
\end{equation}
As  it is mentioned above, the matter effect in $\Delta _{eff}$ can be "eaten"
by the relativistic motion of matter. It follows that the condition for maximal
mixing of $\nu_L$ and $ \nu_R$ in a magnetic field $\vec B$ (which is realised
\cite {ELStpl00} in the vacuum when neutrino is propagating nearly 
perpendicular to the magnetic
field, $ B \approx B_{\perp}, B_{\parallel} \approx 0$) could be realised also
in presence of matter moving with relativistic speed.

From the above discussion on matter effects in neutrino spin oscillations it 
follows that the similar suppression of matter term exist in neutrino 
oscillations without change of helicity for the case of matter  
moving with relativistic speed.

It worth to be noted that if matter is not flavour symmetric (like an 
"ordinary" matter which contains electrons, neutrons and protons, but not 
muons and taons) the matter contribution to the effective 
neutrino potential depends on the neutrino flavour. That is why modification
of matter effects in neutrino oscillations will be different not 
only for different speeds of matter components, but also will be varied 
for neutrino oscillations between various neutrino flavour states. 

In conclusion, we argue that the discussed phenomena of sufficient change of 
matter effects in neutrino oscillations in case of matter motion with 
relativistic speed could have important consequences in astrophysics.

\section*{Acknowledgements}

We thank A.Akhmedov, V.Berezinsky, C.Giunti, V.Naumov, S.Petcov,  
and A.Smirnov for helpful discussions.
We should also like to thank A.Borisov, P.Eminov 
and P.Nikolaev for consultations on the relativistic quantum statistics.

\end{document}